\documentclass[12pt, letterpaper]{article}
\usepackage[top=1in, bottom=1.5in, left=1in, right=1in]{geometry}
\usepackage[utf8]{inputenc}
\usepackage{booktabs}
\usepackage{hyperref}
\usepackage{graphicx}
\usepackage{caption}
\usepackage{gensymb}

\usepackage{natbib}

\newcommand{\simsmaf}{{\tt sims\_MAF}~}
\newcommand{\opsim}{{\tt OpSim}~}

\title{The Definitive Map of the Galactic bulge}
\author{Oscar A. Gonzalez, Will Clarkson, Victor P. Debattista, \\Christian I. Johnson, R. Michael Rich,\\ Giuseppe Bono, Massimo Dall'Ora, John Gizis, Nitya Kallivayalil, \\Daisuke Kawata, Phil Lucas, Dante Minniti, Ricardo Schiavon,\\ Jay Strader, Rachel Street, Elena Valenti, \& Manuela Zoccali}
\date{Nov 2018}

\begin{document}

\maketitle

\begin{abstract}
\noindent 
We recommend configuring the LSST coverage of the inner Galactic plane to allow the production of the definitive age/metallicity map of the Galactic bulge from LSST data, matched to external surveys where appropriate. This will allow the formation history of the Galactic bulge to be reconstructed, as well as furnishing a huge legacy dataset to support one of the key LSST science goals (Mapping the Milky Way). We recommend precise multi-color $grizy$~photometry as deeply as spatial crowding will allow (ideally completing early in the project), complemented by a single-filter survey spread over the entire 10-year time baseline. Both strands should cover as broad an area within the Bulge as possible, with image quality sufficient to reach at least the bulge main sequence turn-off in seeing-limited observations. We specify metrics and a figure of merit by which candidate observing strategies could be evaluated with respect to ``static'' bulge science (proper motions and photometry).
\end{abstract}

\section{White Paper Information}
\begin{enumerate} 
\item {\bf Science Category:} Mapping the Milky Way
\item {\bf Survey Type Category:} Wide-Fast-Deep survey 
\item {\bf Observing Strategy Category:} An integrated program with science that hinges on the combination of pointing and detailed observing strategy.
\end{enumerate}  
\clearpage
{\bf 4. Author information:}\\
~\\

Oscar A. Gonzalez (UK Astronomy Technology Centre: \url{oscar.gonzalez@stfc.ac.uk})\\
\hspace*{6mm}Will Clarkson (University of Michigan-Dearborn: \url{wiclarks@umich.edu})\\
\hspace*{6mm}Victor P. Debattista  (UCLAN: \url{VPDebattista@uclan.ac.uk})\\
\hspace*{6mm}Christian I. Johnson (Harvard-Smithsonian CfA: \url{cjohnson@cfa.harvard.edu})\\
\hspace*{6mm}R. Michael Rich (UCLA: \url{rmr@astro.ucla.edu})\\ 
\hspace*{6mm}Giuseppe Bono (Universita' di Roma)\\
\hspace*{6mm}Massimo Dall'Ora (INAF)\\
\hspace*{6mm}John Gizis (University of Delaware)\\ 
\hspace*{6mm}Nitya Kallivayalil (University of Virginia)\\ 
\hspace*{6mm}Daisuke Kawata (University College London)\\ 
\hspace*{6mm}Phil Lucas (University of Hertfordshire)\\ 
\hspace*{6mm}Dante Minniti (Universidad Andr\'{e}s Bello)\\ 
\hspace*{6mm}Ricardo Schiavon (Liverpool-John Moores University)\\ 
\hspace*{6mm}Jay Strader (Michigan State University)\\ 
\hspace*{6mm}Rachel Street (Las Cumbres Observatory)\\ 
\hspace*{6mm}Elena Valenti (European Southern Observatory)\\
\hspace*{6mm}Manuela Zoccali (P. Universidad Cat\'{o}lica))

\clearpage

\section{Scientific Motivation}


Mapping the Milky Way is a key science driver for the LSST project. However, as pointed out by Strader et al. (2018 whitepaper), {\it half} the stellar mass of the Milky Way (MW) lies in the traditional ``zone of avoidance'' for the main Wide-Fast-Deep survey. Increasing the coverage of the inner Plane to Wide-Fast-Deep levels would have only a minor impact on science out of the plane (at the 0.03-0.05 mag level; Section \ref{ss:timereq}), but would revolutionize the Galactic science delivered by the project (as pointed out in a number of complementary whitepaper-proposals, such as Strader et al., Street et al., Lund et al., Dall'Ora et al., Olsen \& Szkody, and Prisinzano \& Magrini). Here we add a recommendation that proper sampling of the inner Plane be configured in order to permit uncovering the mysterious formation and evolution history of the Galactic bulge, via proper motions and integrated photometry.

~\\
\noindent{\bf 2.1. Why the Galactic bulge is important:} Sitting at the centers of the potentials of galaxies, bulges are important actors in the formation and evolution of spiral galaxies, but their origins and evolution remain hotly debated. One set of mechanisms invokes the merging of sub-units at high redshift (the so-called ``classical bulges,'' or alternatively the merging of gas-rich disks; e.g. \citealt{immeli2004, elmegreen2008}) while the other invokes dynamical instabilities of bars, leading to boxy/peanut-shaped (B/PS) bulges \citep[e.g.][]{combes81, athanassoula2005}. More than 75\% of barred galaxies in the MW's mass regime show B/PS bulges \citep{erwinDebattista2017}, as does the MW itself (see \citealt{zoccaliValenti2016} for a review).

Because it can be studied on a star-by-star basis, the MW bulge offers a unique opportunity to probe the relative formation pathways of galaxy bulges and their hosts. Although we have some confidence in asserting that the evolution of the MW bulge was dominated by dynamical instabilities of the bar, there is evidence that its stellar population is further complicated spatially by age, abundance, and kinematics (see \citealt{barbuy2018} for a review). A number of investigations have addressed these questions, but {\it LSST offers the best opportunity to answer them by producing a sample of high quality kinematics, ages, and metallicities for millions of stars}. The dynamical instability formation mechanism predicts different spatial and kinematic distributions for stars of different ages and metallicity (\citealt{debattista2017}; see also Figure \ref{fig:ages}). Conversely, a bulge with a significant merger component would primarily differ in the inner $\sim$5 degrees, where the clumps come to rest, and along the minor axis, where in the secular scenario the stars are levitated by the bar’s instabilities. {\it The large scale distribution of relative ages therefore probes the formation history of the MW’s bulge} - and thus that of the majority population of spiral galaxies in the mass regime of the MW.

~\\
{\bf 2.2. The confusing age distribution of the MW bulge:} While a small intermediate-age population has been uncovered
\citep[e.g.][]{trapp2018, deguchi2000}, the age distribution of the majority bulge population remains controversial. Most studies of the Main Sequence turnoff (MSTO) in the bulge colour-magnitude diagram
(CMD) suggest that the bulk of bulge stars are $\sim$10 Gyr old with at most a few percent ($<$3\%) of young stars
\citep[e.g.][]{ortolani1995, kr2002, zoccali2003, clarkson2008, valenti2013}. The comprehensive recent analysis by \citet{renzini2018} of the four small fields of the HST WFC3 Bulge Treasury Program \citep{brown2009} finds that metal-rich and metal-poor stars have a consistent luminosity function at the MSTO, hence appear homogeneously $\sim$10 Gyr old, with at most a few percent of stars as young as $\sim$5Gyr. On the other side of the debate, however, \citet{haywood2016} concluded - also from CMD data - that a wide range of ages in the bulge is required to simultaneously reproduce the narrow width of the observed MSTO and the known spectroscopic metallicity of the same field. Deepening the puzzle, spectroscopic indications from $\sim 90$~microlensed main sequence stars near the MSTO suggest roughly 20-25\% of objects are in the $<$ 5Gy range \citep{bensby2017}. As observational studies become increasingly sophisticated, then, the true age distribution in the bulge appears to be growing {\it less} settled.

~\\
{\bf 2.3. Why LSST is needed:} All the studies above are hampered by limited coverage of the sky, whether due to the small field of view traditionally suggested for high-precision photometry and proper motion to decontaminate the bulge from foreground objects (e.g. with HST), or due to strong selection effects for certain tracers (such as the microlensed dwarfs of \citealt{bensby2017}). However, several groups are showing that seeing-limited wide-field imagers are capable of useful photometry at least as deep as the MSTO: at least five DECam bulge projects are ongoing, with the most relevant to the proposed observations being the Blanco DECam Bulge Survey of \citet{rich2017}. With its huge survey grasp and high specified precision, a survey of the inner Galactic plane with LSST would allow the MW bulge to be {\it homogenously} probed over large enough length-scales that the spatial distributions of its constituent tracer populations can finally be mapped.  Main measurements:
~\\
\hspace*{6mm}{\bf 2.3.1. Integrated photometry:} An LSST multi-band photometric survey at a 5$\sigma$~depth fainter than the level of the MSTO on the cleaned CMD of the bulge can be used to determine photometric stellar population indices (similarly to those from \citealt{brown2009}). We assume that the MSTO in lower latitude fields will be difficult to reach in u-band, even for LSST. However, coupling a 5-band survey (grizy) with JHKs photometry from the Vista Variables in the Via Lactea (VVV; \citealt{minniti2010}) survey and its eXtension (VVVX) would compensate for the lack of u-band depth (see for example \citealt{casagrande2019} for $T_{\rm eff}$, [Fe/H] calibrations), resulting in an impressive 9-band dataset covering from the optical to
the near-IR with a $\sim$25y survey time-baseline. 

{\bf 2.3.2. Proper motions:} LSST should be capable of sufficient proper motion precision ($\sim$0.3 mas/yr) to separate bulge stars from the foreground disk, down to the MSTO over most of the bulge. With the capability to sample
fields that are too crowded for Gaia, LSST will be able to smoothly extend the Gaia astrometric error curve down to the turn-off even in highly crowded regions (and for $|b|>$ 5$\degree$~or so, Gaia motions themselves can be used to improve the LSST proper motions; \citealt{ivezic2012,rich2017}).

These measurements would produce {\it global} constraints on the formation and evolution mechanisms at work in MW-like bulges. It would also yield an enormous, homogenous dataset for legacy studies - not just of the bulge itself but also of the disk and inner halo.

\begin{minipage}[t]{7.5cm}
    \vspace{0pt}
        \includegraphics[width=7.5cm]{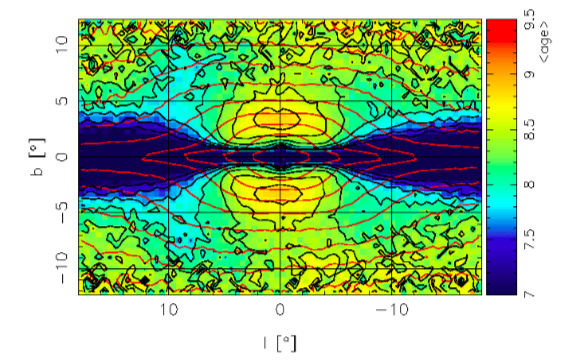}
    \captionof{figure}{The average age map in a model of a galaxy with a pure
      boxy/peanut bulge, comparable to the Milky Way, from
      \citet{debattista2017}.  The average age is highest on the minor
      axis, while away from the axis a range of ages are found. Our proposed LSST survey would enable us to produce the {\it observational} counterpart of this figure.}
    \label{fig:ages}
\end{minipage}
\hspace*{3mm}
\begin{minipage}[t]{8cm}
\vspace{0pt}
\includegraphics[width=8cm]{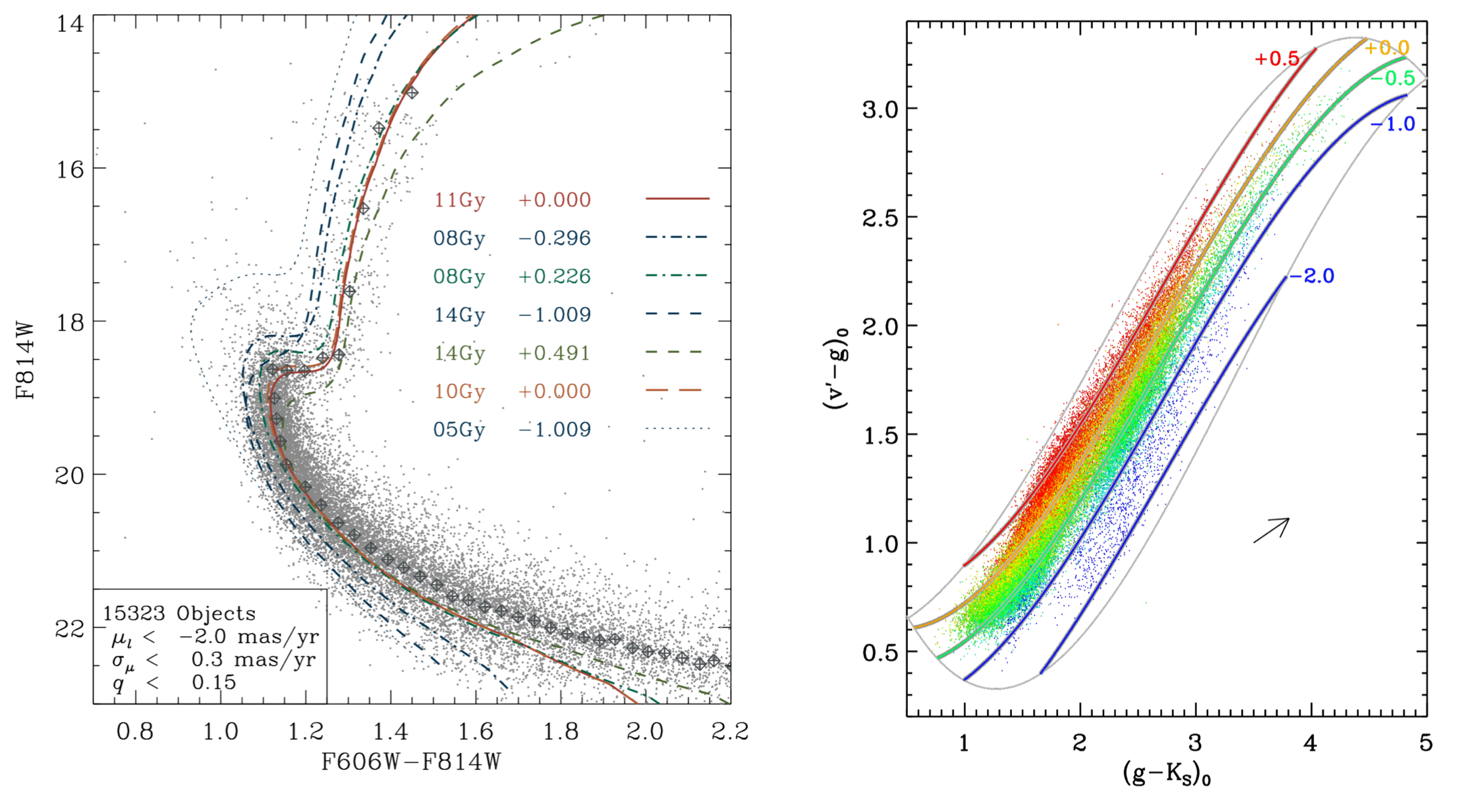}
\captionof{figure}{Left: The MSTO age distribution of a proper-motion clean
      bulge sample in SWEEPS field from \citet{clarkson2008}. Right:
      Example of a colour-colour calibration, colour-coded by
      spectroscopic metallicities, from \citet{casagrande2019}. Such
      photometric calibrations can be used to break the
      age-metallicity degeneracy at the MSTO level.}
      \label{fig:calibrations}
\end{minipage}

~\\
\noindent
\begin{minipage}[t]{16.5cm}
\includegraphics[width=16cm]{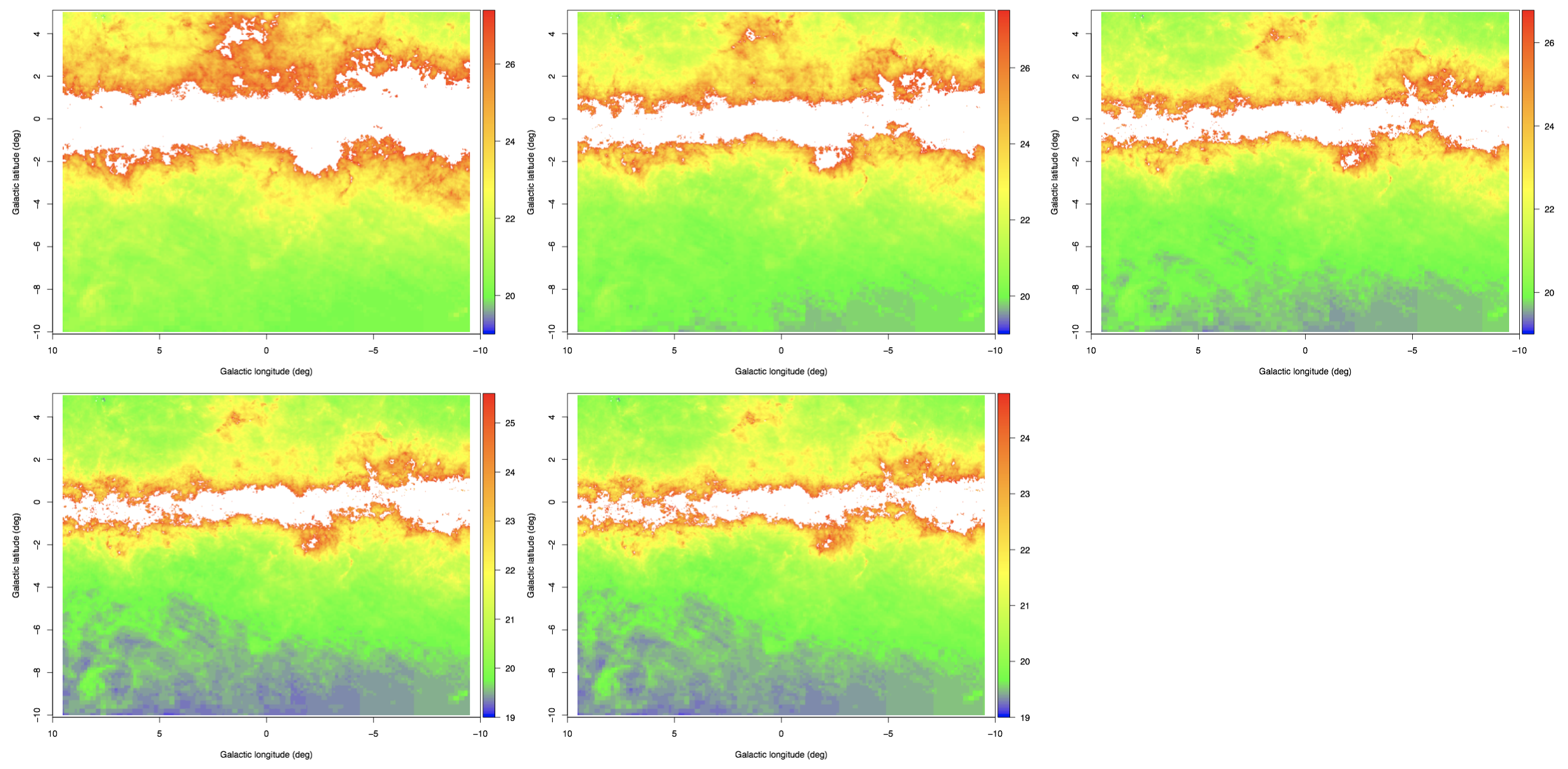}
    \captionof{figure}{Mean magnitude (grizy, reading left-right and top-bottom) of the bulge Main-Sequence
      faintest population (10 Gyr, Solar metallicity) in the bulge
      region covered by the VVV survey (note that VVVX completes the
      $5\degree<b<10\degree$ gap in VVV) accounting for reddening
      \citep{gonzalez2012} and bar position angle of 27$\degree$. The white
      areas correspond to the regions where the MSTO magnitude is
      fainter than the co-added depth in {\tt
        astro\_lsst\_01\_1004\_n128\_t9999}. Panels show $(-10\degree < l < +10\degree)$~and $(-10\degree < b < +5\degree)$.}
    \label{fig:spatial}
\end{minipage}

\vspace{.6in}

\section{Technical Description}

\subsection{High-level description}
\label{ss:highlev}
We suggest partitioning Wide-Fast-Deep-like coverage of the inner Plane into two sub-surveys, each of which is required to achieve the main science objective:

\begin{itemize}
    \item[{\bf S1:}]{A single-band (i-band) multi-epoch sub-survey, with observations spread over the 10 years time-baseline down to i=26.8 mag;}
    \item[{\bf S2:}]{a multi-colour sub-survey (S2) mapping the same region as S1 in as many bands as possible (griyz) with 5-sigma depth fainter than the MSTO.}
\end{itemize}
The MSTO of the Galactic bulge changes in magnitude according to the
amount of reddening reaching the faintest magnitudes at (grizy) =
(27.5, 27.3, 26.8, 26.0, 24.5) for latitudes $|b|>2\degree$. S1 will be used
for proper-motions $<$ 0.3 mas/yr (or less when coupled with
VVV/DECam) which will be used to 'clean' the CMDs obtained with S2
from the disc population so that it can be used to determine
population indices with the combination of LSST+VVV filters of the
(pure) bulge population at the level of the MSTO and brighter.

~\\
(Note that the absence of u-band photometry in our description does not mean that we are advocating LSST should avoid observing the plane in u-band - but merely that we do not rely on u-band photometry for the science communicated in this proposal. Of course, u-band exposures would greatly benefit a variety of science cases in addition to our own; see, e.g. the cadence whitepapers of Strader et al., Dall'Ora et al. and Street et al.)


\subsection{Footprint -- pointings, regions and/or constraints}

The survey requires as complete spatial coverage as possible between
$-15\degree < l < +15\degree$ and $-10\degree < b < +10$\degree. Pointings should be the same for S1
and S2.

\subsection{Image quality}

Given the level of crowding of the Galactic bulge region at the level
of the MSTO, seeing conditions better than 0.8" are required.

\subsection{Individual image depth and/or sky brightness}

S1 requires multiple visits in only one band, with individual depth of
each visit to the MSTO level, spread over the 10 years to improve the
accuracy of proper-motions. The individual depth is not particularly
relevant for S2, provided that co-added depth is achieved in all
bands.

\subsection{Co-added image depth and/or total number of visits}

Multi-band observations are required for S2, in as many bands as
possible with resulting co-added depth fainter than the MSTO (grizy) =
(27.5, 27.3, 26.8, 26.0, 24.5).

Our \opsim tests based on {\tt astro\_lsst\_01\_1004\_n128\_t9999}
indicate that the 5$\sigma$~magnitude depth can be reached within the
current LSST scheduling across the entire bulge region outside of the
area closest to the mid-plane ($|b|>2\degree$, see
Fig.~\ref{fig:spatial}). We note that \opsim and \simsmaf predictions
including spatial confusion are somewhat pessimistic compared to
experience with DECam, a 4m seeing-limited {\it ugrizY} imager that in
many ways is the pathfinder instrument for LSST. For example, \opsim and \simsmaf predict that the crowding limit for LSST will be above
the MSTO for the low-reddening Baade’s Window (at l,b=1,-4, \simsmaf
suggests limiting magnitude y$\sim$19 at 5$\sigma$). However, several
groups are reaching below the MSTO even in these crowded regions using
a handful of DECam exposures per filter per field. The limiting magnitudes
achieved in practice with DECam pathfinder surveys are not far below
the limiting magnitudes suggested by \opsim \& \simsmaf without
considering crowding (typically $\sim$1-2 magnitudes brighter, varying
by field, but still fainter than the MSTO in most regions for $|b| >
2\degree$). We expect the existing \simsmaf~confusion metric will be made
more realistic in light of DECam experience, and so for the present
have opted not to include the \simsmaf confusion metric in our
analysis for this proposal.

\subsection{Number of visits within a night}

We do not have specific constraints on the number of visit for
individual nights.

\subsection{Distribution of visits over time}

S1 requires epochs distributed over the entire 10y survey. S2 would
strongly benefit from completion of the multi-colour observations
during the first year to activate as many early science cases as
possible. Such a strategy has proven successful by the excellent
scientific outcome of the VVV survey by scheduling multi-colour
observations during the first year and subsequent 10-years of
multi-epoch Ks imaging. In addition, early completion of the deep observations for S2 would aid programs in time-domain science that require multi-color observations to aid with object classification (e.g. the Plane survey whitepapers by Street et al., Lund et al., Strader et al. and Dall'Ora et al.).

\subsection{Filter choice}

S1 requires multi-epoch observations in i-band. S2 requires
multi-colour observations in as many bands as possible from the LSST
filter-set (except u-band).

We do not include u-band in this proposal as it would be difficult to match the co-added depth requirements of other bands across the entire survey area. However, we note that shallower u-band coverage would be very valuable for classification of brighter objects and additional metallicity estimates from RGB stars.

\subsection{Exposure constraints}

The number of exposures in a visit should be sufficient for cosmic ray
removal, and to permit sufficient dynamic range to reach the bulge MSTO as well as constrain stars on the bulge Giant Branch.

While measurements at the standard $2\times 15$~s~visit exposure time would fully satisfy our requirements at the level of the MSTO, we note that the RGB tip (which would provide additional information for us as well as activating other science cases) would saturate at the standard exposure time. We therefore suggest the possibility of having (5s+5s) exposure pairs in addition to the standard (15s+15s) pair, or possibly a revised visit sequence consisting of two pairs of unequal exposure duration (e.g. two pairs per visit at 15s+5s per pair). We anticipate iterating with the LSST Project to determine how best to incorporate unequal-exposure pairs into metrics and figure of merit estimates.

\subsection{Other constraints}

No additional constraints.

\subsection{Estimated time requirement}
\label{ss:timereq}

In this proposal we are advocating a particular distribution of exposures across the ten-year survey time baseline; sufficient depth in the first year to reach the MSTO in even the more reddened regions, in {grzy} filters (this is the S1 requirement), and a distribution across the entire decade-long baseline in i-band to measure proper motions (S2). Assuming the LSST survey is configured to well-sample the inner Bulge - the simplest version of which would be to extend the Wide-Fast-Deep survey to the inner plane, as advocated in the Strader et al. and Street et al. whitepapers, among others - then our proposal adds {\bf no additional time requirement} to the survey. 

With the understanding that the time required for our science would also enable the wide range of inner-Galactic science (see, e.g. the whitepapers of Strader et al., Street et al., Bono et al., Lund et al., Olsen \& Skody, and  Prisinzano \& Magrini), we can make a rough estimate of the time required for this science by applying the nominal exposure specifications under a Wide-Fast-Deep-like cadence, covering the survey region of the bulge, which should be roughly correct even allowing for details such as the incorporation of short exposures. Assuming 40s per visit, WFD-like visit totals per filter per field, considering only exposures in {grizy}, and assuming that at least 2/3 of the bulge is so covered, we estimate a total of roughly 1000 hours to perform WFD-like coverage over the entire inner bulge region. 

A more detailed estimate of the acceptable total time will be provided by the metrics and figure of merit that we specify in this proposal. We point out that a previous assessment of the extension of Wide-Fast-Deep to the inner Plane suggests that only very minor impact to extragalactic science would result (the faint limit in extragalactic fields being impacted only at the 0.03-0.05 mag level, see, e.g. \url{https://github.com/LSSTScienceCollaborations/ObservingStrategy/issues/355}).

\vspace{.3in}

\begin{table}[ht]
    \centering
    \begin{tabular}{l|l|l|l}
        \toprule
        Properties & Importance \hspace{.3in} \\
        \midrule
        Image quality &  1   \\
        Sky brightness &  3 \\
        Individual image depth &  2 \\
        Co-added image depth &  1 \\
        Number of exposures in a visit   &  3 \\
        Number of visits (in a night)  &  3 \\ 
        Total number of visits &  1 \\
        Time between visits (in a night) &  3\\
        Time between visits (between nights)  &  3 \\
        Long-term gaps between visits & 1 \\
        Other (please add other constraints as needed) & \\
        \bottomrule
    \end{tabular}
    \caption{{\bf Constraint Rankings:} Summary of the relative
      importance of various survey strategy
      constraints. Considerations are given a rank, from 1=very
      important, 2=somewhat important, 3=not important.}
        \label{tab:obs_constraints}
\end{table}

\subsection{Technical trades}

Both spatial coverage and survey depth are critical for this
proposal. However, co-added depth to the level of the MSTO is
absolutely required for each observed field. In the case of a
trade-off between area coverage and field depth, the multi-band
co-added depth in every observed field should be prioritized. Area
coverage is fundamental for reconstruction of large-scale trends of
the bulge stellar populations with kinematics and
morphology. Therefore, if survey area is compromised, this proposal
would benefit from fields well spread over the requested area ($-15\degree < l
< +15\degree$ and $-10\degree < b < +10\degree$) rather than sampling a small knot of contiguous fields. The proposal
does not place any constraints on number of visits or the timings of
those visits beyond the spacing of multi-epoch observations across the
10yr survey.

\section{Performance Evaluation}

Requirements S1 and S2 (see Section \ref{ss:highlev}) are both essential for static science. The metric and figure of merit should thus heavily penalize fields that do not satisfy both S1 and S2 (fields covered in two filters only are not useful for this science). In addition, spatial coverage is important; assuming that fully complete coverage may be difficult, a candidate observing strategy that adequately observes a high spatial dispersion of fields is strongly preferred. These considerations find expression in the metrics and figure of merit on the following page.
~\\
\clearpage
\noindent {\bf Metrics} for LSST Bulge static science:
\begin{enumerate}
    \item {{\it Depth1g, Depth1r, Depth1i, Depth1z, Depth1y:} Formal limiting magnitude at 5$\sigma$~photometric precision after one year of operation, including crowding but updated to account for observational experience towards the inner Plane;}
    \item{{\it PM10y}: Formal proper motion precision at the apparent magnitude of the bulge main sequence turn-off (the latter could be supplied by lookup table; e.g. Figure \ref{fig:spatial}.)}
\end{enumerate}

\noindent{\bf Figure of merit} for LSST Bulge static science: We suggest $FoM \equiv f/f_{\rm ideal}$, where
\begin{equation}
    f = N_{\rm good}s_l s_b \label{eq:fom}
\end{equation}
and $f_{\rm ideal}$~represents (\ref{eq:fom}) evaluated for the hypothetical survey with complete coverage of all fields in the survey region. In (\ref{eq:fom}), $N_{\rm good}$~is the number of fields for which the following compound condition holds:
\begin{itemize}
    \item ($Depth1g > MSTOg$ + 3 mag) \& 
    ($Depth1r > MSTOr$ + 3 mag) \& \\
($Depth1i > MSTOi$ + 3 mag) \& 
($Depth1z > MSTOz$ + 3 mag) \&\\ 
($Depth1y > MSTOy$ + 3 mag) \& ($PM10y < \mu_{\rm max}$)
\end{itemize}
with $\mu_{\rm max}$~the proper motion precision required (which varies with Galactic latitude depending on the existence and capability of existing proper motion catalogs in these regions). The terms $s_l$~and $s_b$~are the rms of the field centers in Galactic longitude and latitude (respectively) of only the fields that meet the compound condition above. 


\section{Special Data Processing}

\begin{figure}[h]
\centering
\includegraphics[width=8cm]{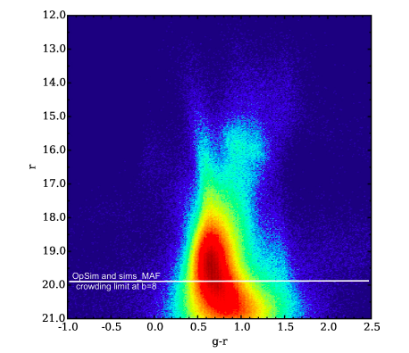}
\caption{Typical color-magnitude diagram towards the inner bulge, from
 a handful of DECam images taken as part of the Blanco DECam Bulge
 Survey (BDBS; PI R. Michael Rich, photometric processing by
 C. I. Johnson). The MSTO is clearly visible, with the LSST \simsmaf
 suggested faint limit indicated as a horizontal white line. Even in
 this relatively uncrowded region, DECam is producing useful
 photometry to a somewhat greater depth than currently suggested by
 LSST \simsmaf performance metrics. We anticipate refinement of the
 LSST metrics using DECam experience will improve the accuracy of the
 LSST metrics.}
\label{fig:cmd}
\end{figure}

Tests based on DECam data (see for example Fig.\ref{fig:cmd}) suggest
that an approach based around the industry-standard {\tt DAOPHOT}
routines for crowded-field photometry will likely out-perform the
pipelines used for the wider LSST survey (bulge regions are far more
spatially crowded than most of the main-WFD survey fields, and
extended-source photometry is not a major requirement for bulge
fields). At this date we anticipate running {\tt DAOPHOT}-based
software pipelines on LSST bulge data.

We expect to request the use of community time on the LSST processing
hardware to run these pipelines. Based on experience with massively
multi-core architecture when run on DECam data, we anticipate these
pipelines will be readily adapted to run on LSST processing
architecture. Should the processing needs exceed what is reasonably
available through LSST-supported hardware, we would seek institutional partnerships and external funding to support this
processing and calibration. The DECam experience (e.g. with the Blanco
DECam Bulge Survey, P. I. R. Michael Rich, which uses photometric
pipelines developed by Christian I. Johnson) suggests that institutional-level
resources are appropriate to producing large, calibrated photometric
datasets as long as real-time delivery is not required (i.e. when
processing time on the order of weeks-months does not negatively
impact the utility of the data). To ensure project success, we intend
to seek external salary funding at the level of one FTE for at least
two years, to support a postdoc-level individual to manage the data
processing, calibration, and delivery to the community.

We do anticipate ultimately using the LSST Science Platform to host
the ~5 billion object catalog (aggregated positions, magnitudes,
proper motions, and associated uncertainties) that our mini-survey
will generate. These products will resemble a subset of the "Data
Release level" (formerly Level-2) standard products, and we expect to
iterate with the DM team to determine the best methods to store and
host these products.
\clearpage
\section{References}
\begin{small}
\begingroup
\bibliographystyle{aasjournal}
\setlength{\bibsep}{3pt}
\bibliography{bulgeStatic}
\endgroup
\end{small}
\clearpage
\pagebreak

\end{document}